\documentclass{article}
\usepackage[utf8]{inputenc}

\usepackage{geometry}
\usepackage{graphicx}
\graphicspath{ {./images/} }
\usepackage{hyperref}

\title{Josephson dynamical simulation using the electronic circuit simulator APLAC: a tutorial}
\author{Mikko Kiviranta}
\date{}

\begin{document}
\maketitle

\begin{abstract}Analysis Program for Linear Active Circuits (APLAC) is a general-purpose electronic circuit simulator, which has included a built-in model of the Josephson junction (JJ) since late 80's. It capabilities in simulating eg. noisy Superconducting Quantum Interference Devices (SQUIDs), Rapid Single Flux Quantum (RSFQ) logic circuits, or superconducting Transition Edge Sensors (TESes) are relatively unknown within the superconducting electronics community. Here we give a brief step-to-step tutorial for APLAC users to unleash those capabilities.
\end{abstract}

\section{Introduction}
APLAC is a general-purpose electronic circuit simulator whose humble beginnings \cite{APLAC2, APLACtwente}, are almost as old as the much more widely-known SPICE simulator. Josephson junction was available already in its early versions, at least from the version 6.24 which the author learned to use in the early 90's. Such functionality has been useful and continues to be useful in designing Josephson devices, including SQUIDs and SQUID arrays with parasitics. APLAC is a contemporary of a number of other JJ-equipped simulators including the WR-SPICE \cite{WRSpice} and JSIM \cite{JSIM}, but less well known.
 
A lot of effort went into development of APLAC after Nokia Corporation began using it as their major microwave design tool. The NASSE schematic editor was developed along with the originally text-based APLAC simulation engine. The APLAC version 6.24 (maybe already earlier) included the integrated schematic capture, which feature made the APLAC attractive for Josephson dynamical simulations involving parasitics. Our earlier, hand-written code (eg. \cite{TapaninSimu,MikonSimu}) had to be re-written and re-compiled whenever circuit topology (i.e. schematics) changed. The 6.24 was a 16-bit application in Windows, so that memory limitations prevented long simulation runs. From the version 7.10 onwards the simulator was a 32-bit Windows application, wich together with the rapidly increasing processor speeds made Josephson simulations feasible even in this kind of an intepreted rather than compiled form. 

The spin-off company APLAC Inc was established to sell and further develop the simulator as stand-alone software, but in 2005 the company was acquired by Applied Wave Research and APLAC got merged into their Microwave Office suite. More recently, APLAC has been acquired along with the AWR by the Cadence Design Systems.

This document is intended as a companion to the paper 'Superconductive circuits and the general-purpose electronic simulator APLAC' accepted to {\sl IEEE Transactions on Applied Superconductivity}, and to act as a wrapper to the relevant simulation code files. This tutorial was originally located as \href{http://virtual.vtt.fi/virtual/SQUID/APLAC/}{a web page}.

\subsection{Josephson junction model using controlled sources: LT-SPICE}
It is straighforward to use a controlled current source to realize the first Josephson relation, \(I(t) = I_{C} \: sin(2\pi\:\theta(t))\) and to re-intepret a voltage of an internal node to present the quantum phase \(\theta(t)=\frac{1}{\Phi_{0}}\:\int_{0}^{t}U(\tau)\:d\tau\). Quantum phase can then be generated by an integrator driven by the instantaneous voltage across the J-junction. Most straighforward an integrator is a controlled current source charging a capacitor. 

\begin{figure} \centering
\includegraphics[width=7.25cm]{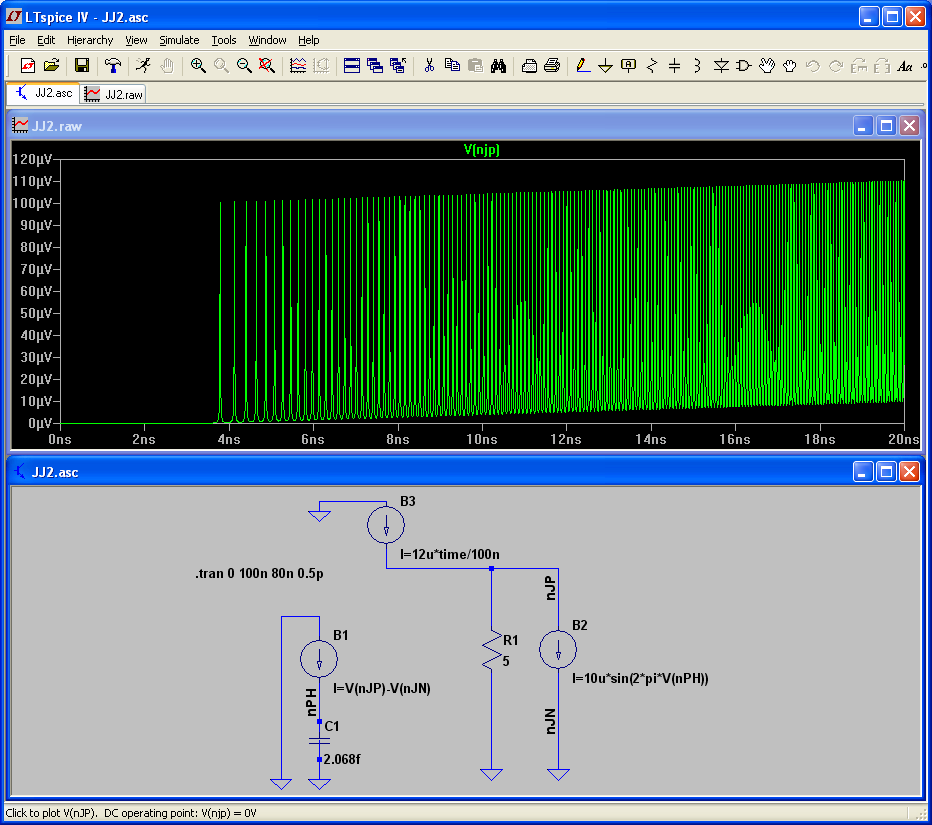} \includegraphics[width=7.25cm]{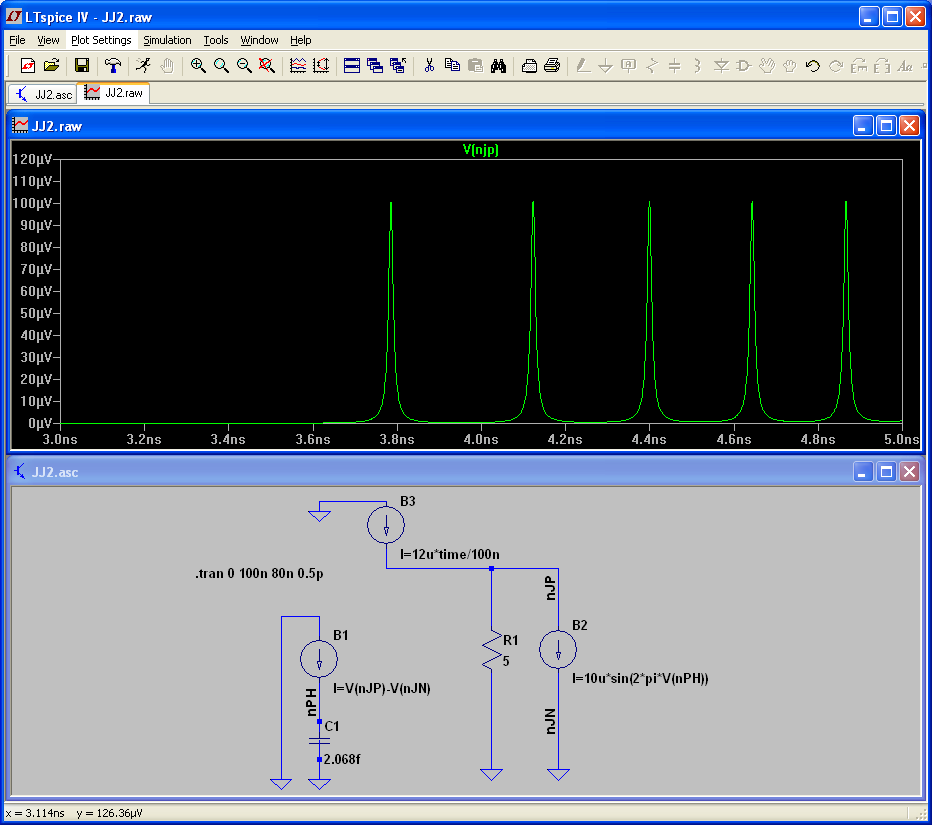}
\caption{Josephson junction modelled by LT-SPICE IV} \label{JJ_ltspice}
\end{figure}

Shown in Fig.\ref{JJ_ltspice} is a resistively shunted J-junction implemented in the \href{https://en.wikipedia.org/wiki/LTspice}{LT-SPICE} version IV, with the schematic {\tt JJ2.asc} available as an ancilliary file. The J-junction model with \(I_{C}=100\mu A\) is driven here by a bias source ramping the current up to \(J_{C}=0\ldots 120\mu A\) in 100 nanoseconds. The {\tt .tran} directs recording of the final 20 nanoseconds of voltage across {\tt B2}, which corresponds the \(I_{B}=96\mu A\ldots 120\mu A\). The record shows the onset and increasing frequency of the Josephson oscillation as the bias current exceeds \(I_{C}\). 

\begin{figure} \centering
\includegraphics[width=12cm]{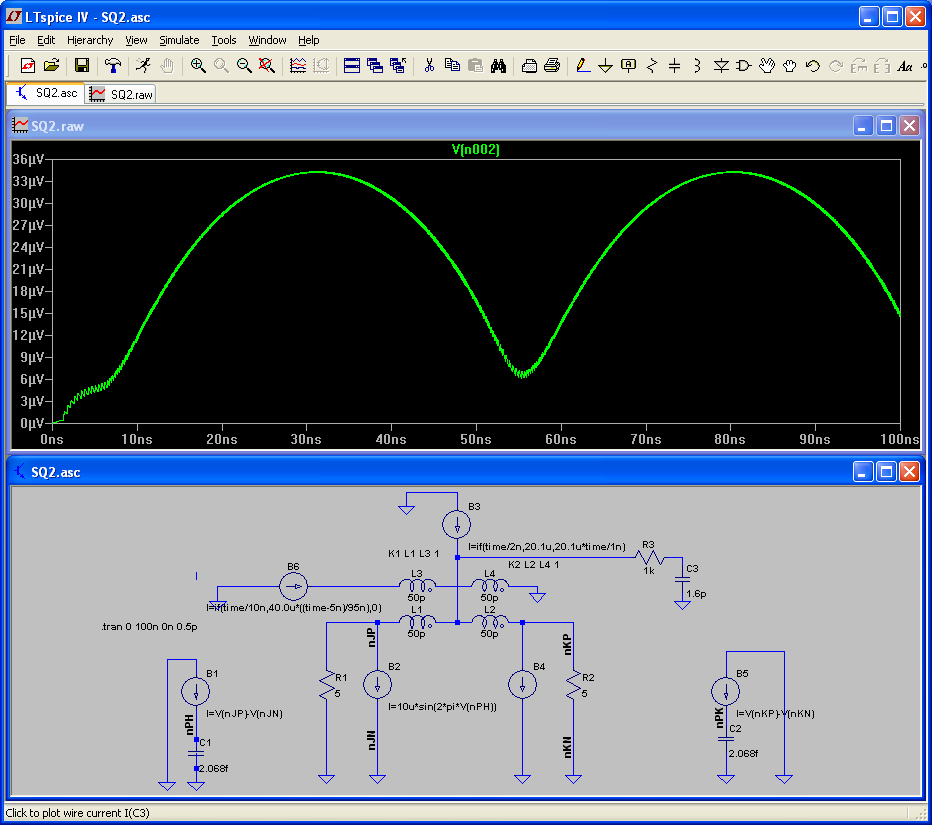}
\caption{DC-SQUID modelled by LT-SPICE IV} \label{SQ_ltspice}
\end{figure}

A full dc-SQUID can be constructed from such J-junctions as shown in Fig.\ref{SQ_ltspice} (LT-SPICE schematic {\tt SQ2.asc} available as ancilliary). In the simulation the first 2ns are used to ramp up the SQUID bias current \(I_{B}=0\ldots 20.1\mu A\) , so that \(I_{B}> 2I_{C}\) makes the SQUID remain in the finite-voltage state at all flux values. The applied flux is driven over 2 flux quanta, \(2\Phi_{0}\), during the 100ns total simulation time. The low-pass filter R3/C3 averages out the Josephson oscillation, so that the flux-to-voltage response of the SQUID gets plotted as the voltage across C3. 

\subsection{Josephson junction model using controlled sources: APLAC}
\begin{figure} \centering
\includegraphics[width=7.25cm]{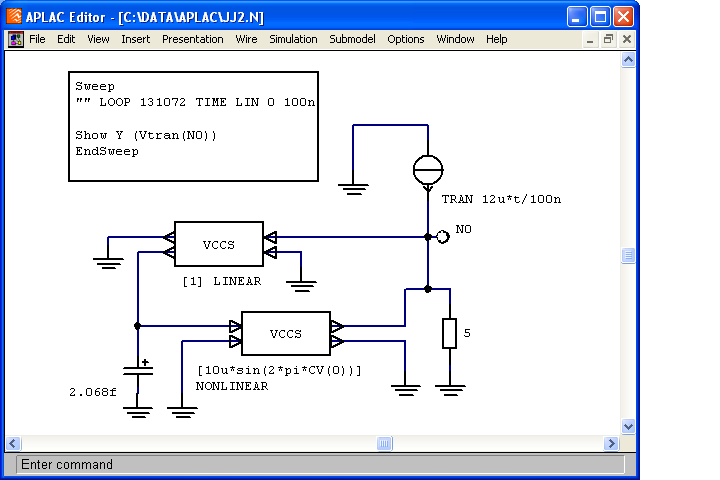}
\includegraphics[width=7.25cm]{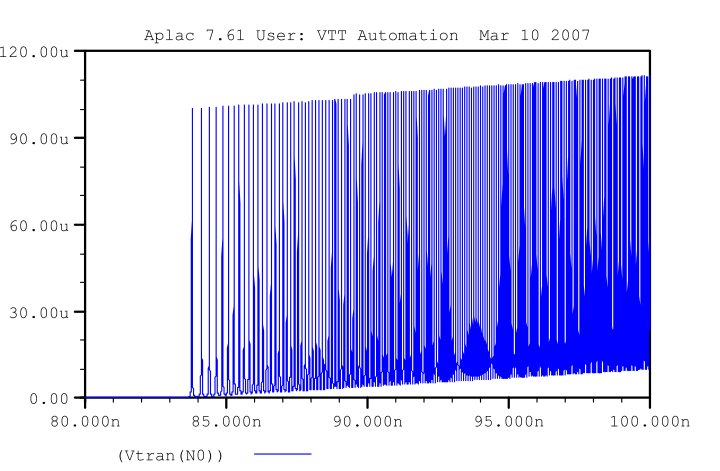}  \includegraphics[width=7.25cm]{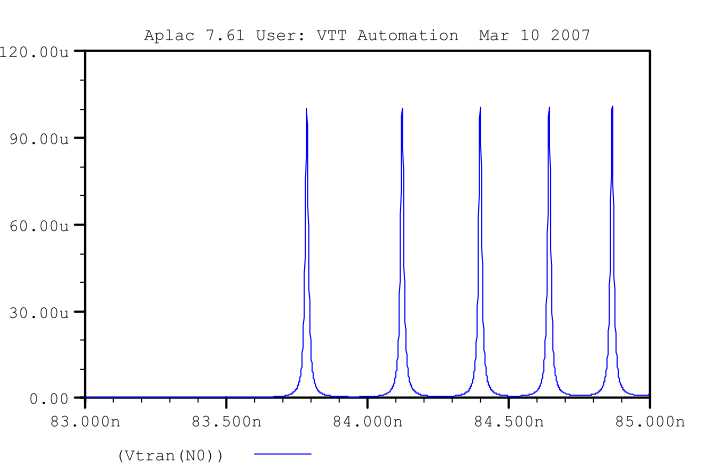}
\caption{Josephson junction modelled by APLAC controlled sources} \label{JJso_aplac}
\end{figure}

With APLAC, the same circuits can be constructed from voltage-controlled current sources (VCCS's) as shown in Fig.\ref{JJso_aplac}. Simulating from the schematic {\tt JJ2.N} and netlist {\tt JJ2.I}, available as ancilliary files, similar time behaviour follows.

\begin{figure} \centering
\includegraphics[width=7.25cm]{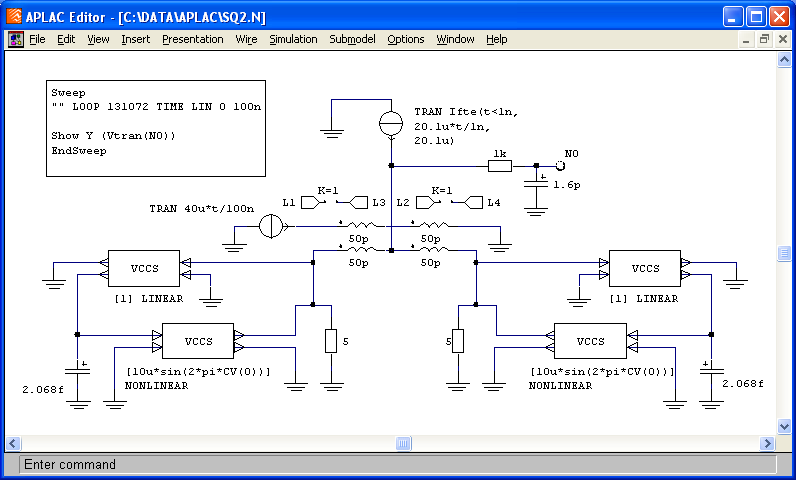}  \includegraphics[width=7.25cm]{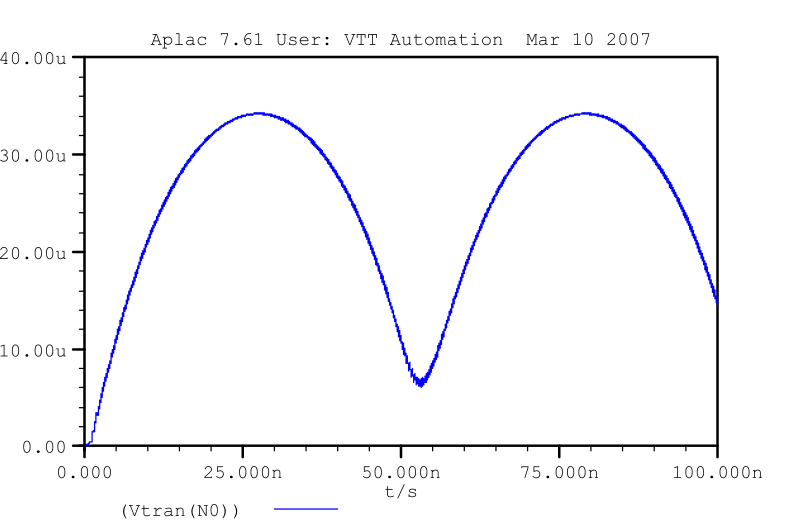}
\caption{DC-SQUID modelled by APLAC controlled sources} \label{SQso_aplac}
\end{figure}

The APLAC version of the dc-SQUID is described in the schematic {\tt SQ2.N} and the {\tt SQ2.I} netlist, available as ancilliary files. The simulation result is shown in Fig. \ref{SQso_aplac}.

\begin{figure} \centering
\includegraphics[width=7.25cm]{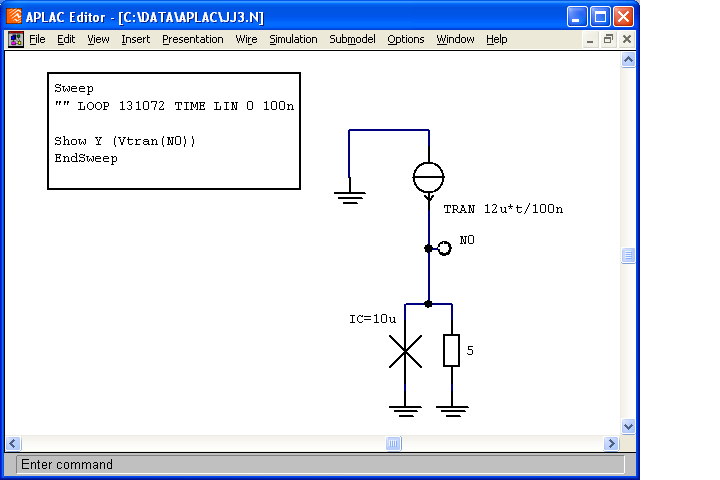}  \includegraphics[width=7.25cm]{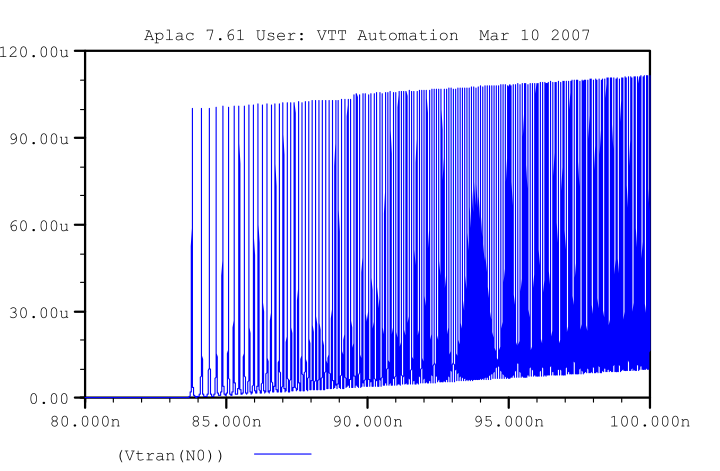}
\caption{Simulating with the built-in JJ in APLAC} \label{JJjj_aplac}
\end{figure}

In APLAC, however, there exists the Josephson junction as a built-in library element. The circuits are much more convenient to construct and simulate faster with the library element. The basic J-junction can be demonstrated with the schematic {\tt JJ3.N} and netlist {\tt JJ3.I} available as ancilliary. The result is the same as by using controlled sources, as seen in the Fig.\ref{JJjj_aplac}.

\begin{figure} \centering
\includegraphics[width=7.25cm]{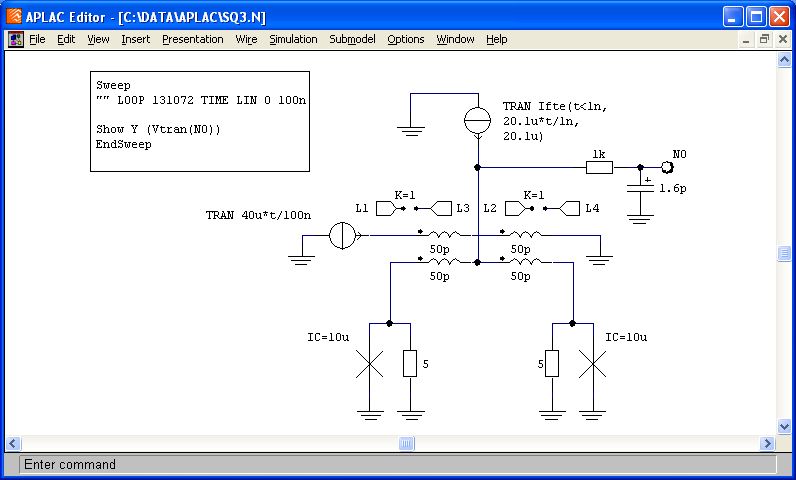}  \includegraphics[width=7.25cm]{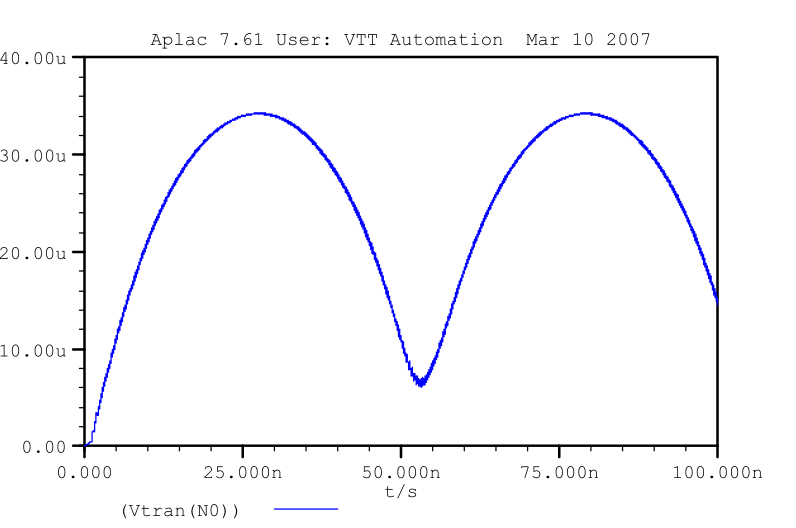}
\caption{DC-SQUID constructed from built-in JJs in APLAC} \label{SQjj_aplac}
\end{figure}

Similarly, the DC-SQUID construction becomes simpler (schematic {\tt SQ3.N} and netlist {\tt SQ3.I} as ancillary), see Fig. \ref{SQjj_aplac}.

\section{Noisy SQUID simulated in APLAC}

 It is straightforward to build a time domain simulation of a dc-SQUID using the built-in JJ element. In superconducting circuits the Kirchoffs voltage law is modified: not only voltages around any closed loop vanish, but also time integrals of voltages (i.e. fluxes) around any closed loop vanish. (More precisely, they do not vanish but are multiples of the flux quantum \(\Phi_{0}\) - this is enforced in simulation if there is a J-junction breaking the loop). To make the APLAC calculate the initial condition correctly, the simulation must start at zero initial currents and voltages, and must be driven into the setpoint explicitly. 

Often, SQUID simulations are performed in dimensionless variables. Here we use SI units for voltages and currents, and realistic values for SQUID parameters. After a simulation run is performed with a given set of SI dimensions, the dimensionless variables \cite{Tesche} tell how the results scale into some other set of SI dimensions. There is an additional quirk that the flux quantum in APLAC is exactly 2.07 femto volt-seconds, not its true value of 2.0678... femto volt-seconds. 

\begin{figure} \centering
\includegraphics[width=11cm]{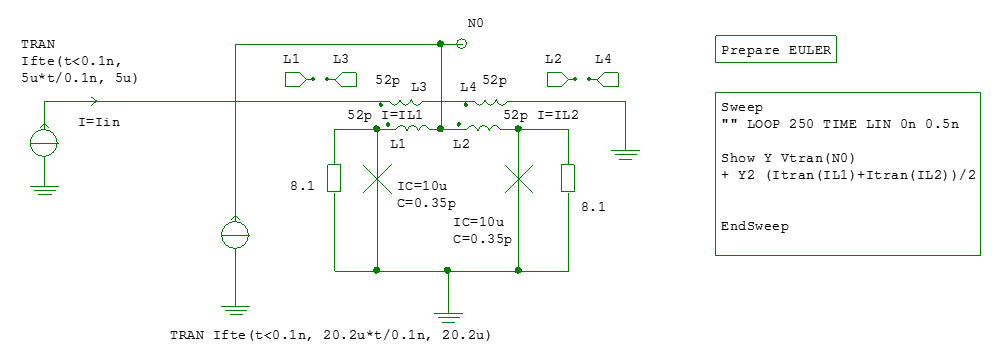}  \includegraphics[width=11cm]{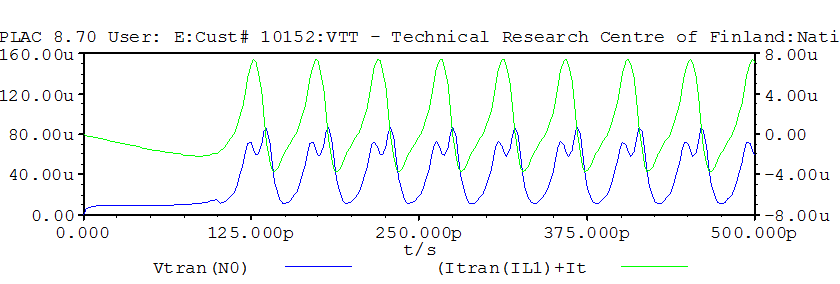}
\caption{The voltage across a DC-SQUID and the circulating current.} \label{SQ_aplac}
\end{figure}

As an example, Fig,\ref{SQ_aplac} shows a simple simulation of the circulating current and the output voltage of a current-biased dc-SQUID with \(\beta_{C}=0.7\), \(\beta_{L}=1.0\), with the applied flux of \(\Phi_{0}/4\) and bias current 101\%  of the SQUID critical current \(2I_{C}\). (We always use \(I_{C}\) to refer to single-JJ critical current. We also use slightly nonstandardly the symbol \( \beta_{L}=\frac{2L_{SQ} I_{C}}{\Phi_{0}} \) for the dimensionless loop inductance). The APLAC schematic {\tt DCSQUID\_A.N} and netlist {\tt DCSQUID\_A.I} files are found in the {\tt /anc} subfolder, and the result in Fig. \ref{SQ_aplac}. 

The circuit contains magnetic mutual coupling between coils L1 and L3, and between L2 and L4. The {\tt Prepare} statement chooses the Euler method for numeric integration, which works better with noisy Josephson junctions than the default Trapezoidal method. The bias current is ramped up to the \(20.2\mu A\) value during the first 0.1 ns of simulation by using the If-Then-Else numerical function within the current source definition. Mutual inductance M = 104 pH from the input coil L3+L4 to the SQUID loop L1+L2 implies response periodicity of \(20\mu A\), so that \(5\mu A\) current ramped up during the first 0.1 ns implies \(\Phi_{0}/4\) applied flux during the simulation. APLAC components are not completely ideal however, specifically the smallest resistance which can occur anywhere is \(10\mu \Omega\) as default, if not changed in the {\tt Prepare} statement. This implies eg. that the circulating current injected to the SQUID loop during the first 0.1 ns will decay with time constant of 52 pH/\(10\mu \Omega\ = 5\mu s\). Therefore, the longest simulation time should be significantly shorter than the ‘supercurrent’ decay time. 

In the above plot, the blue trace is the ~22 GHz Josephson oscillating voltage, measured from node {\tt N0} and referred to left-hand Y-axis. The green trace is the circulating current in the SQUID loop, referred to the right-hand Y-axis. 

\subsection{Filtering to plot SQUID characteristics}

\begin{figure} \centering
\includegraphics[width=8.5cm]{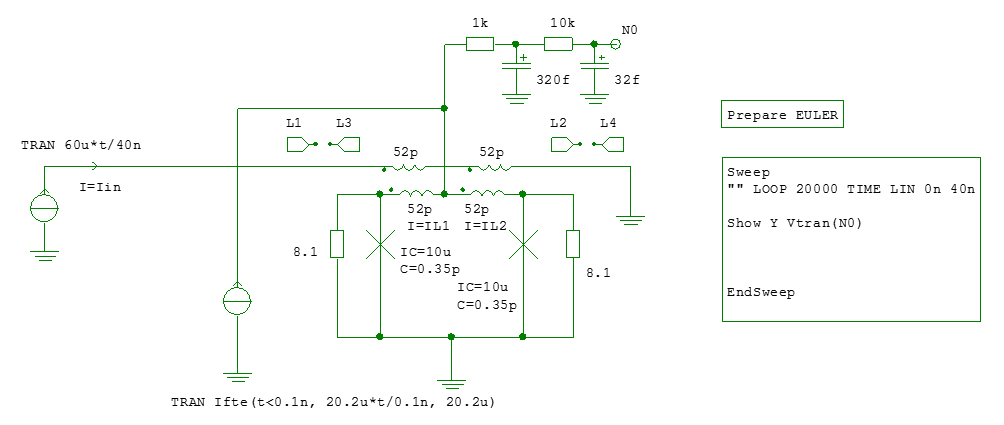}  \includegraphics[width=6cm]{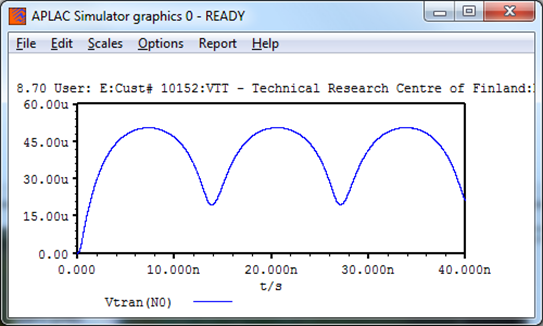}
\caption{Use of filtering to abtain DC-SQUID characteristics.} \label{SQ_filt_chars}
\end{figure}

When generating SQUID characteristics, it is necessary to filter away the Josephson oscillation and only retain the average voltage. We implement a 2-pole low-pass filter with a 0.5 GHz corner frequency to the circuit, utilizing the ideal nature of the components which allows use of ridiculously low or high values. The corner frequency corresponds to the Josephson frequency at \(1\mu V\) and should hence work well whenever the voltage across SQUID is larger than a few \(\mu V\). Due to settling time of the filter, we must increase the total simulation time. In the example below, the input coil current is ramped over \(60\mu A\) i.e. three flux quanta during the 40 ns simulated time. The APLAC schematic {\tt DCSQUID\_B.N} and netlist {\tt DCSQUID\_B.I} are found as ancillary files, results in Fig. \ref{SQ_filt_chars}. 

\begin{figure} \centering
\includegraphics[width=8.5cm]{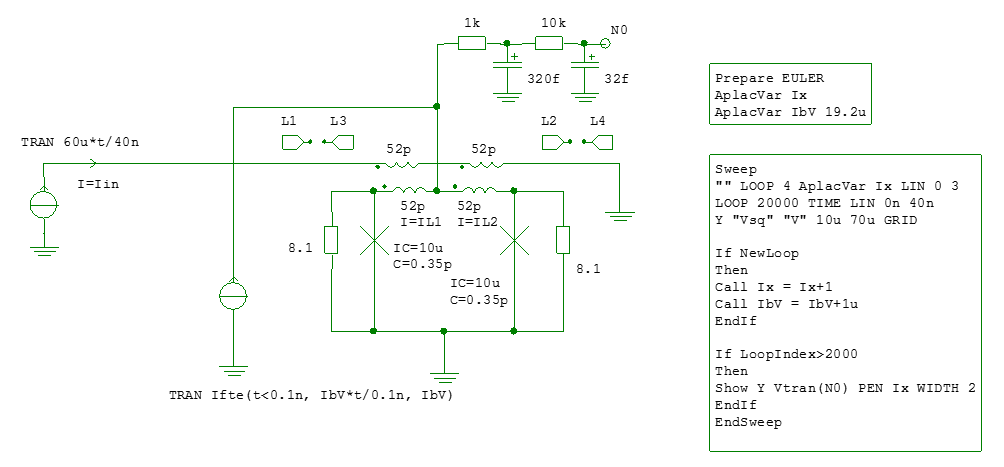}  \includegraphics[width=6cm]{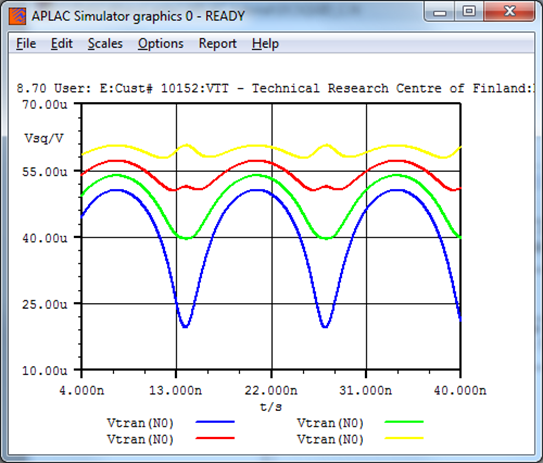}
\caption{DC-SQUID flux characteristics at several bias current values.} \label{SQ_chars_manycurr}
\end{figure}

A set of flux characteristics can be plotted by introducing variables {\tt Ix} and {\tt IbV} by the {\tt AplacVar} statement and by adding another nested loop to the {\tt Sweep}. The schematic is {\tt DCSQUID\_C.N} and netlist {\tt DCSQUID\_C.I}. The code generates flux-to-voltage characteristics at bias currents \(I_{B}=20.2, 21.2, 22.2 \) and \( 23.2\: \mu A\), see Fig. \ref{SQ_chars_manycurr}. 

\begin{figure} \centering
\includegraphics[width=8.5cm]{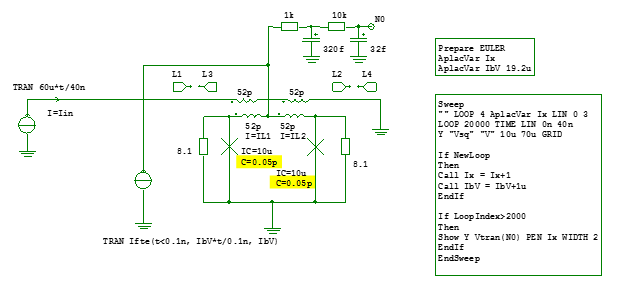}  \includegraphics[width=6cm]{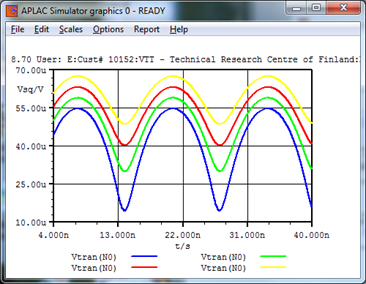}
\caption{DC-SQUID flux characteristics, LC crossing moved to higher voltage.} \label{SQ_chars_noLC}
\end{figure}

The responses show the well known IV-curve crossing at half-integer applied flux, due to the resonance of the loop inductance \(L_{SQ}\) and junction capacitances \(C_{J}\). It is easy to verify that the crossing disappears if the junction capacitance (hence \(\beta_{C}\)) is lowered, see Fig, \ref{SQ_chars_noLC}. 

\subsection{Introducing noise}

\begin{figure} \centering
\includegraphics[width=11cm]{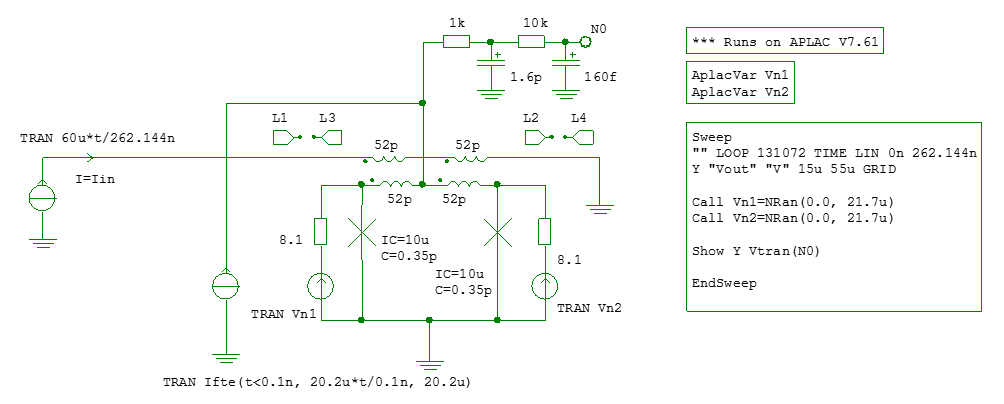}
\caption{DC-SQUID with explicit noise sources.} \label{SQ_expl_noise}
\end{figure}

To introduce noise to the circuit, we originally inserted voltage sources representing the Johnson noise in series with the shunt resistors. Noting that voltage noise density of \(8.1\, \Omega \) resistors is \(43\, pV\, /\, Hz^{1/2}\) at T = 4.2 K, whose RMS value is \(21.7\, \mu V\) over the 250 GHz Nyquist band\footnote{Assuming brickwall frequency response} implied by the 2 ps time step, we included gaussian random number generators with zero mean and standard deviation of \(21.7\, \mu V\). With this approach which we used with the APLAC version 7.61 it was crucial to store the instantaneous noise voltage values into variables at each time step. This guaranteed that (i) the Nyquist bandwidth is correct, and (ii) the instantaneous noise voltage didn't change during the intra-timestep adaptation of the APLAC numerical engine. The relevant schematic file is {\tt DCSQUID\_D\_761.N} and netlist file is {\tt DCSQUID\_D\_761.I}, see Fig.\ref{SQ_expl_noise}. 

\begin{figure} \centering
\includegraphics[width=8.5cm]{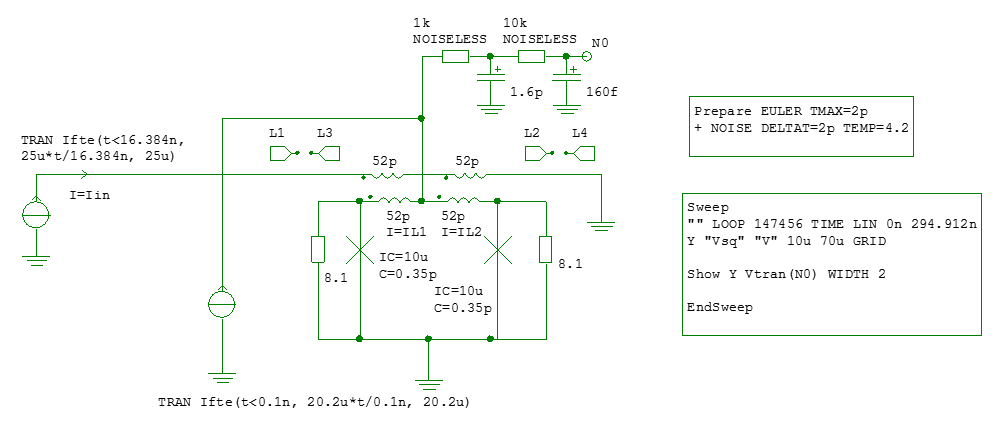} \includegraphics[width=6cm]{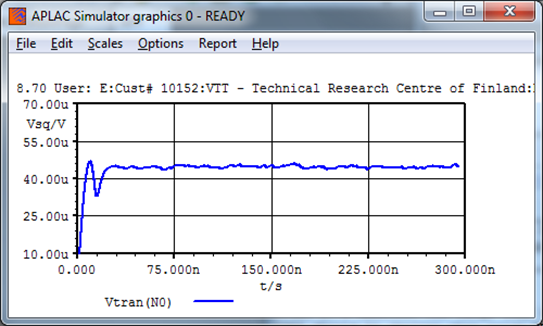} 
\caption{DC-SQUID with internal noise sources.} \label{SQ_int_noise}
\end{figure}

The more modern version APLAC 8.70 no longer executes the above construct, but rather, one must trust the APLACs internal noise models. In the example of Fig.\ref{SQ_int_noise}, we have chosen \(2^{14}\) –step (32.768 ns) settling phase followed by a \(2^{17}\) –step (262.144 ns) noise collecting phase. The initial settling period allows a lower 0.1 GHz corner frequency for the averaging filter. We first sweep the flux over 1.25 periods in order to get a visual clue of the periodic response for debugging purposes, and settle at \(I_{B} = 20.2 \mu A\) bias, \(\Phi_{0}/4\) applied flux. The added line in the {\tt Prepare} statement selects the time discretation step and temperature of the internal APLAC 8.70 noise sources. The resistors in the averaging filter should be defined as noiseless.

\begin{figure} \centering
\includegraphics[width=6cm]{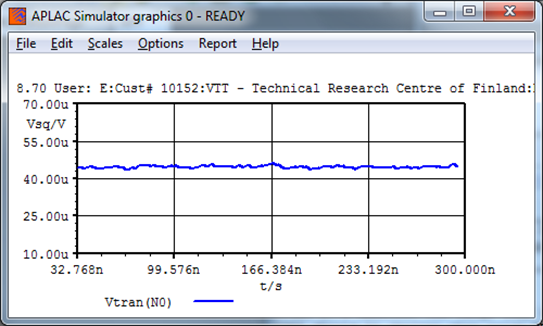} \includegraphics[width=6cm]{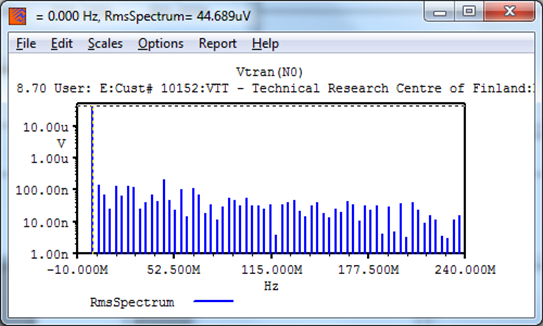}
\caption{(Left) Noisy time trace with the initial transient removed by changing the displayed time start value. (Right) Fourier Transform of the time trace.} \label{SQ_tcropped}
\end{figure}

Now, one can rescale the APLAC timetrace to start only after the settling transient, Fig.\ref{SQ_tcropped} (left). Then from the menus choose {\tt Report, Fourier transform} with the selections {\tt Fast, RMS, Volts} and {\tt Rectangular}. When the lowest voltage in the Y-scale of the resulting spectral plot is chosen non-zero (e.g. 1 nV here), it is possible to choose {\tt Options, Log Y} axis and rescale the X-axis (frequency) conveniently, resulting the Fig.\ref{SQ_tcropped} (right).

The amplitude of the 0 Hz bin, as measured above with the {\tt Options, Probe} menu selection, gives correctly the same value as the dc voltage in the time domain trace. From the Fourier amplitudes the eight bins 3.815, 7.629 … 30.517 MHz, sufficently far below the corner frequency of the averaging filter, the root-mean-square sum is \(99.3\, nV_{RMS}\) or \(51\, pV/Hz^{1/2}\) scaled by the bin width. One could now estimate the steepness \(dV/d\Phi\) of the flux response of the SQUID at the setpoint and calculate the effective flux noise. However, we will automate the procedure in the next example.

\subsection{Neat plotting of the spectrum}

\begin{figure} \centering
\includegraphics[width=8.5cm]{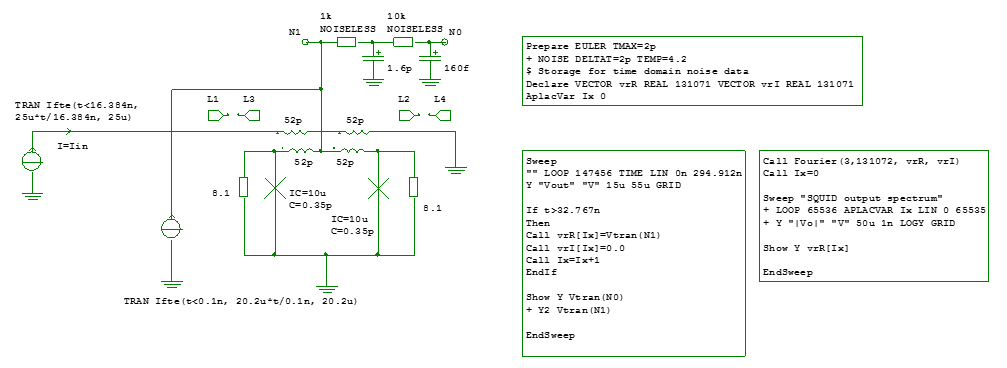}  \includegraphics[width=6cm]{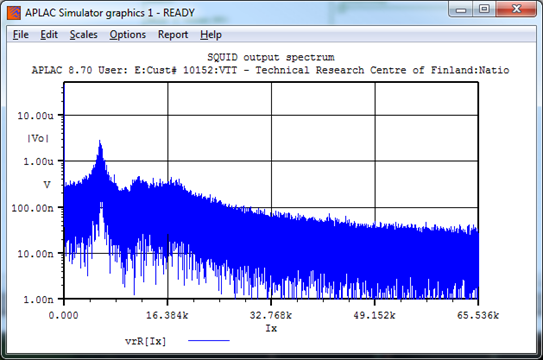}
\caption{Noisy DC-SQUID with the FFT initiated within code.} \label{SQ_fft_in_code}
\end{figure}

First, we will automate the Fourier transform, see Fig. \ref{SQ_fft_in_code}. To avoid the complications associated with the averaging filter, we’ll introduce an unfiltered tap named {\tt N1}. After the settling time has expired, real part of the time domain signal is stored into a vector variable {\tt vrR} and imaginary part equalling zero into {\tt vrI}. The APLAC's built-in function {\tt Fourier} performs the Fourier transform, where the argument {\tt 3} directs the result to be stored stored in magnitude-phase format. In APLAC, the {\tt Fourier(3, xxx)} returns the amplitude \(a_{N}\) of the Fourier term \(a_{N}\: sin(n \omega t + \theta)\), i.e. \(\sqrt{2}\) times the RMS signal falling within the frequency bin. The second Sweep plots the generated spectrum, and allows us to format the plot conveniently. X-axis units are FFT bin numbers. Along the Y-axis there is plotted the RMS voltage noise at the SQUID output, at the one-bin bandwidth. Schematic and netlist files are {\tt DCSQUID\_E.N}, {\tt DCSQUID\_E.I}. 

\begin{figure} \centering
\includegraphics[width=8.5cm]{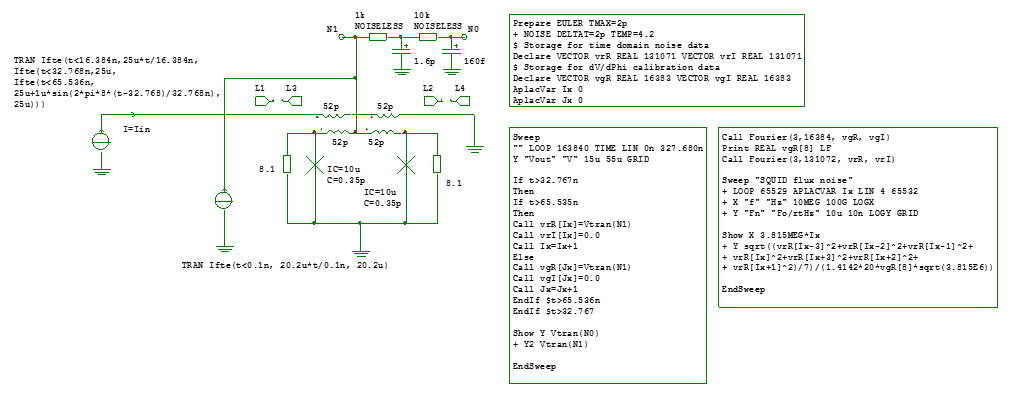}  \includegraphics[width=6cm]{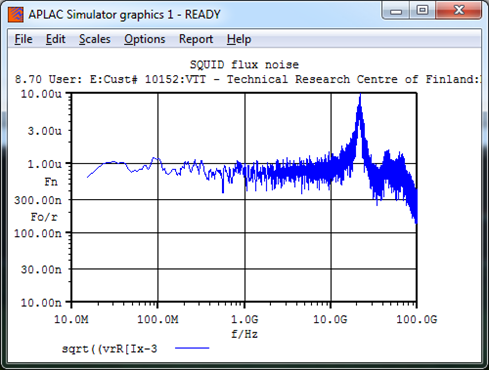}
\caption{Spectral density of the DC-SQUID flux noise.} \label{SQ_fluxnoise}
\end{figure}

In the next simulation there are two \(2^{14}\) -step initialization intervals followed by one \(2^{17}\) -step noise gathering stage. The first interval lets the simulator settle to the chosen bias and flux setpoints. In the next interval, SQUID is driven by the sinusoidal flux excitation of 0.1 \(\Phi_{0\: p-p}\), at frequency which precisely corresponds to the 8th bin of the Fourier transform, for the sake of determining the SQUID gain \(dV/d\Phi\) . The plot can be made neater by performing a 7-point sliding average on the spectral data. Using the knowledge that frequency bins are \(\Delta f\, = \, (N\: \Delta t)^{-1}\) or 3.815 MHz apart, the X- axis can be scaled into the units of hertz and Y-data into spectral density per \(Hz^{1/2}\). Schematic {\tt DCSQUID\_F.N} and netlist {\tt DCSQUID\_F.I} files. Finally the measured voltage noise at the SQUID output is scaled by the \(dV/d\Phi\) into the input-referred flux noise, Fig.\ref{SQ_fluxnoise}.

A problem in the previous simulation file is that noise is active also during the \(dV/d \Phi \) determination stage, which leads to a fluctuating estimate. It is more practical to keep the noise sources inactive during the gain determination and switch them on for the noise-gathering stage. This allows use of a lower than 0.1 \(\Phi_{0\:p-p}\) test excitation but neglects the possible noise rounding of sharp features in the flux characteristics.

\begin{figure} \centering
\includegraphics[width=4.8cm]{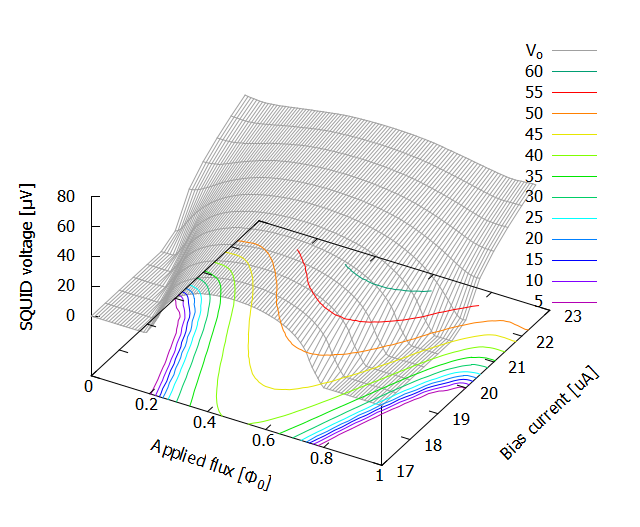}  \includegraphics[width=4.8cm]{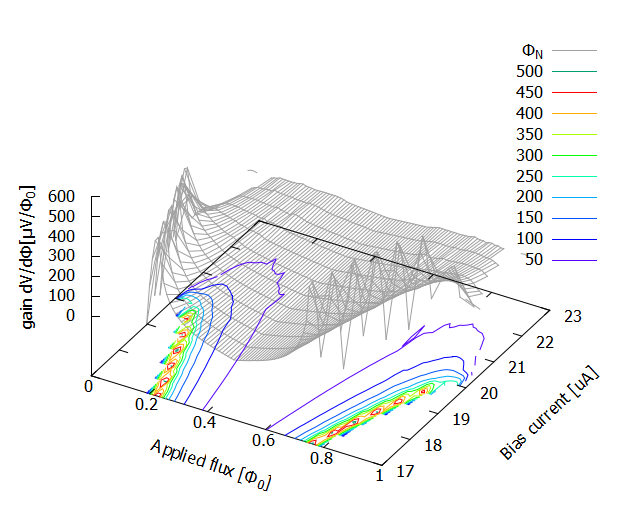} \includegraphics[width=4.8cm]{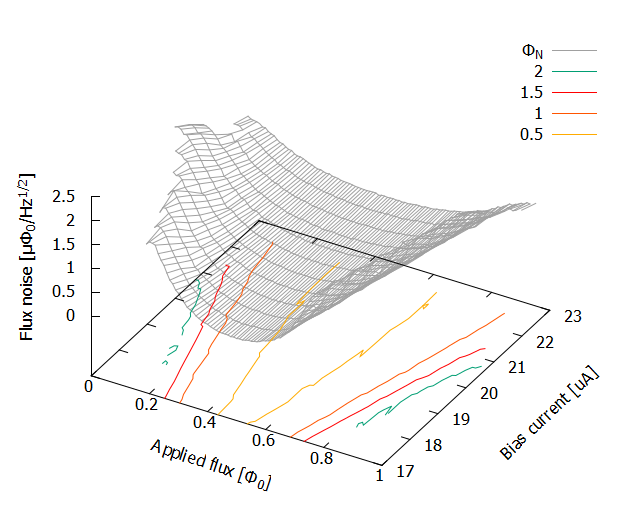}
\caption{Voltage, gain dV/d\(\Phi\) and flux noise of a DC-SQUID as a function of bias current and applied flux.} \label{SQ_surfaces}
\end{figure}

In addition to the output voltage, it is possible to obeserve the circulating current in the SQUID loop, and record its fluctuations to determine the backaction noise. It is also possible to step through a number of SQUID setpoints automatically, and simulate noise behaviour at each setpoint. As an example the result of simulating at 100 flux values and at 13 bias current values is shown in Fig.\ref{SQ_surfaces}. Simulation took 45 minutes on a laptop, when the {\tt Show} statements were disabled to avoid the time used for plotting on the computer screen. The calculated data was stored into files and printed out using GnuPlot.

\section{Simulation of Josephson circuits beyond resistively shunted DC-SQUIDs}

Although our primary interest is the noise behaviour of dc SQUIDs and SQUID arrays, other Josephson junction circuits can be easily simulated, too. 

\subsection{RSFQ logic}
\begin{figure} \centering
\includegraphics[width=7.25cm]{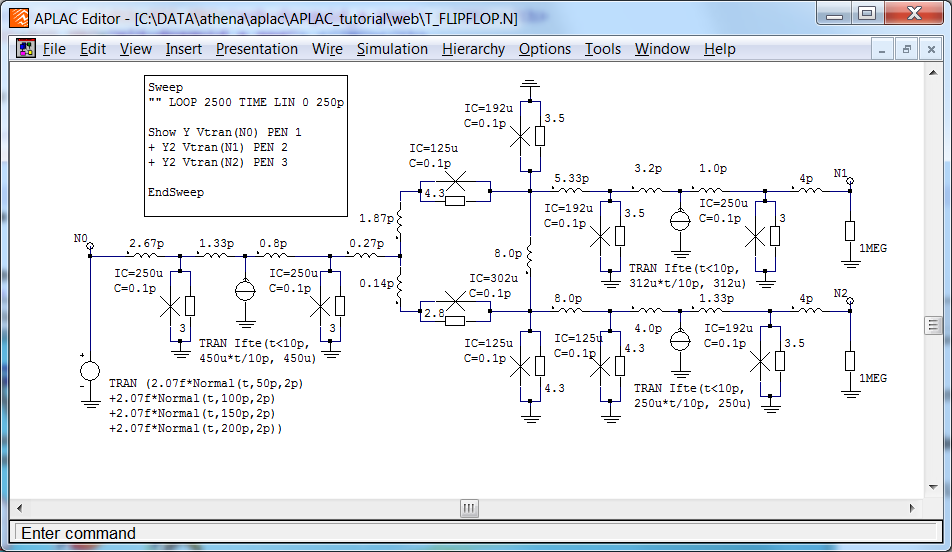}  \includegraphics[width=7.25cm]{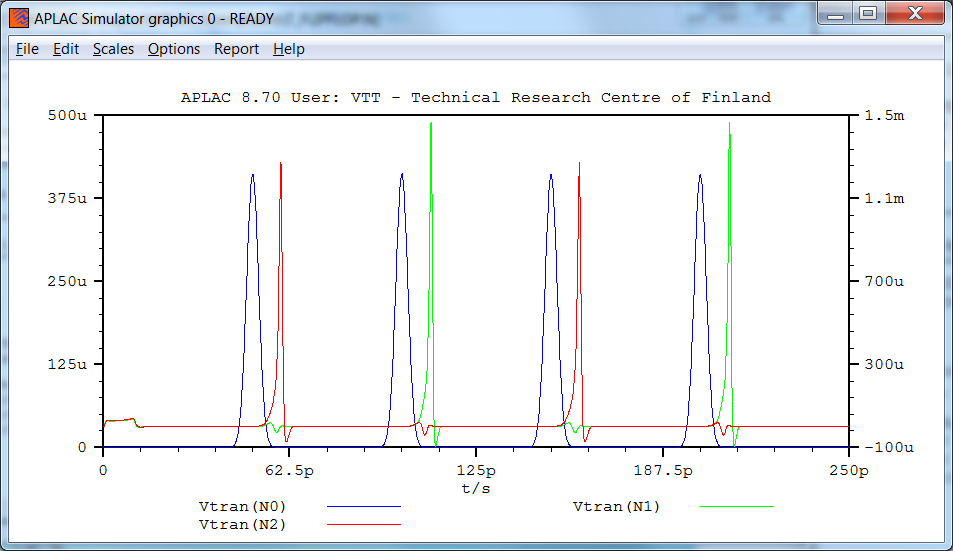}
\caption{RSFQ T flip-flop.} \label{Tflop}
\end{figure}

As an example from the domain of RSFQ logic, Fig.\ref{Tflop} shows simulation of the T-flipflop, taken from \cite{Polonsky}. APLAC schematic and netlist files are {\tt T\_FLIPFLOP.N} and {\tt T\_FLIPFLOP.N}. Blue traces are 2ps wide gaussian RSFQ pulses driving the (T)oggle input, referred to the left-hand vertical axis. Green and red traces are \(Q\) and \(\overline{Q}\) outputs of the flipflop, referred to the right-hand vertical axis. The 10 Josephson junctions are externally shunted to obtain the McCumber-Stewart parameter of roughly \(\beta_{C}=0.7\). 

\subsection{Beyond the RSJ model}
\begin{figure} \centering
\includegraphics[width=7.25cm]{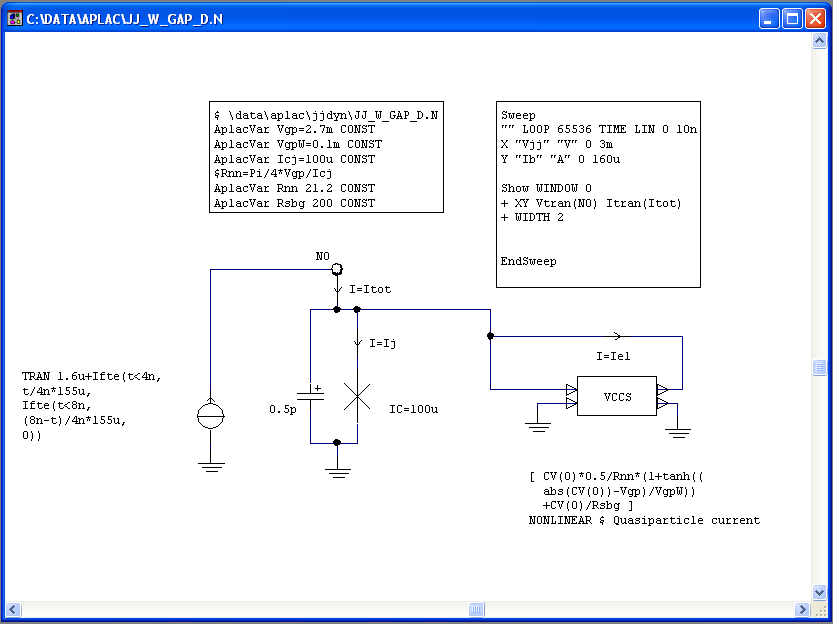}  \includegraphics[width=7.25cm]{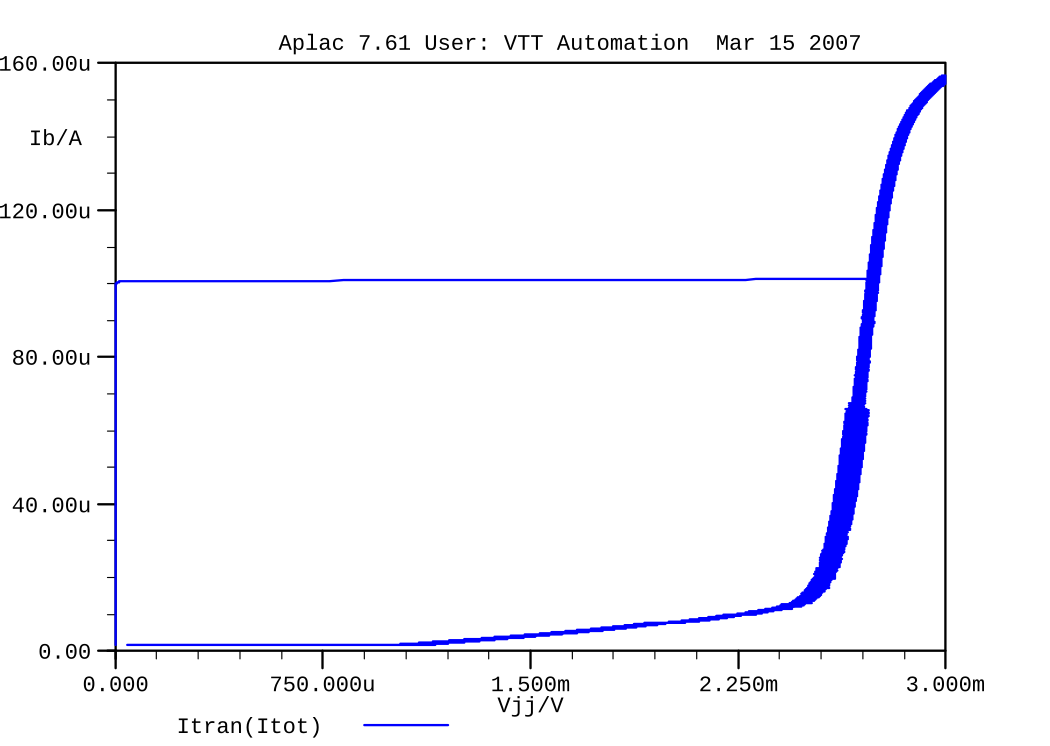}
\includegraphics[width=7.25cm]{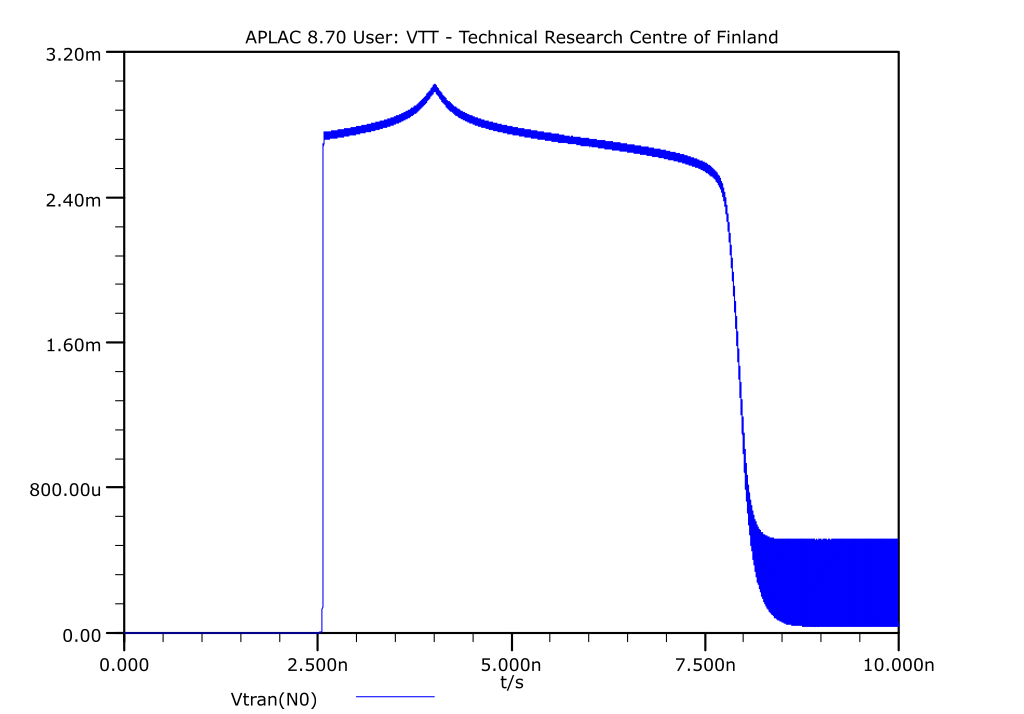}  \includegraphics[width=7.25cm]{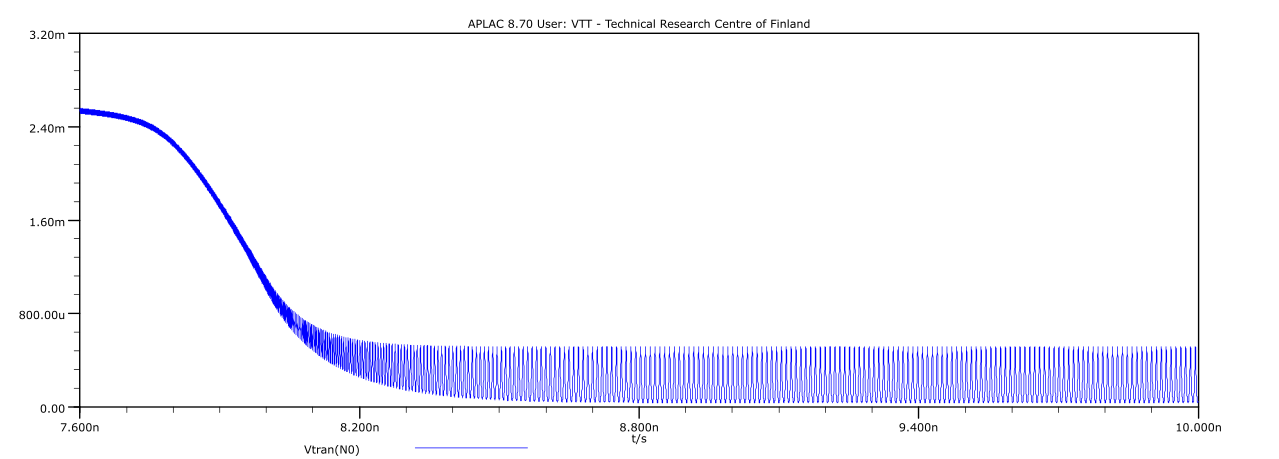}
\caption{Josephson junction with a model for non-linear subgap resistance (top left). Simulated current-voltage characteristics of the junction (top right). Time trace of the simulation where current is ramped up for the first 4 ns and ramped down for the next 4 ns (bottom left). Final 2 ns of the simulation shows how re-trapping to the zero-voltage state does not occur and Josephson oscillation persists (bottom right).} \label{JJ_sbg}
\end{figure}

In all the above circuits the Josephson junctions have been resistively shunted with resistors, with \(\beta_{C}\leq \)1. Although hysteretic JJ circuits can in some cases be simulated without careful modelling of the quasiparticle behaviour, modelling is straighforward by using controlled sources as non-linear resistances. In Fig.\ref{JJ_sbg} the subgap behaviour of a finite-capacitance Josephson junction is modelled with an aid of the hyperbolic tangent function. The adjustable parametes are: {\tt Vgp} the gap voltage, {\tt VgpW} width of the subgap-to-normal transition and {\tt Rsbg} the subgap resistance. The normal-state resistance {\tt Rnn} should be set consistent to its Ambegaokar-Baratoff value. The APLAC {\tt JJ\_W\_GAP\_D.N} schematic and {\tt JJ\_W\_GAP\_D.i} netlist are found among the ancilliary files.

Here the bias current of the JJ is swept from \(I_{B}=1.6\:\mu A\) to \(156.6\:\mu A\) in 4 nanoseconds and back to \(1.6\:\mu A\) in further 4 nanoseconds, motivated by exploring the subgap branch. Finally there is the 2 ns dwell period at \(I_{B}=1.6\:\mu A\). In the upper row there is the junction voltage plotted as a function of the bias current. The voltage-less superconducting branch is followed by the jump to the finite-voltage state at the JJ critical current of \(I_{C}=100\:\mu A\). When the bias current is lowered, the JJ voltage traverses the finite-voltage quasiparticle branch, and remains at a finite voltage at the final \(I_{B}=1.6\:\mu A\). For results see Fig.\ref{JJ_sbg}.

In the lower row there is shown the junction voltage as a function of simulation time. It shows more clearly the ~\(200\:\mu V\) average voltage at the \(I_{B}=1.6\:\mu A\) stationary end state, which is slightly above the retrap voltage \(V_{rtp}\) for these particular junction model values. Due to this the junction remains in finite-voltage state indefinitely. If we choose a slightly lower end current \(I_{B}=1.5\:\mu A\) for the sweep, the JJ decays to the zero-voltage state within the 2 ns dwell time. There is a magnified plot of the final 2.4 ns of the simulation, where the ~100 GHz Josephson oscillation is more clearly visible. 

Above the retrap voltage \(V_{rtp}\) the supercurrent (Josephson) oscillation is so fast that the shunt capacitance keeps the average voltage across the junction a constant, whereby the quantum phase \(\theta\) determined by the Josephson relation \(\frac{d}{dt}\theta=\frac{2\pi}{\Phi_{0}}\langle U \rangle \) proceeds uniformly. In this case the supercurrent \(I(t)=I_{C}\:\sin(\theta)\) averages to zero, so that only the quasiparticle current contributes. 

Below \(V_{rtp}\) the instantaneous voltage across the junction varies during the Josephson cycle, and the supercurrent \(I(t)=I_{C}\:\sin(\theta)\) does not average to zero. Contribution of the supercurrent drives the average voltage of the system further downwards, until the system reaches a zero-voltage state (even at a finite bias current).

\section{Simulating Transition Edge Sensors}

\begin{figure} \centering
\includegraphics[width=7.25cm]{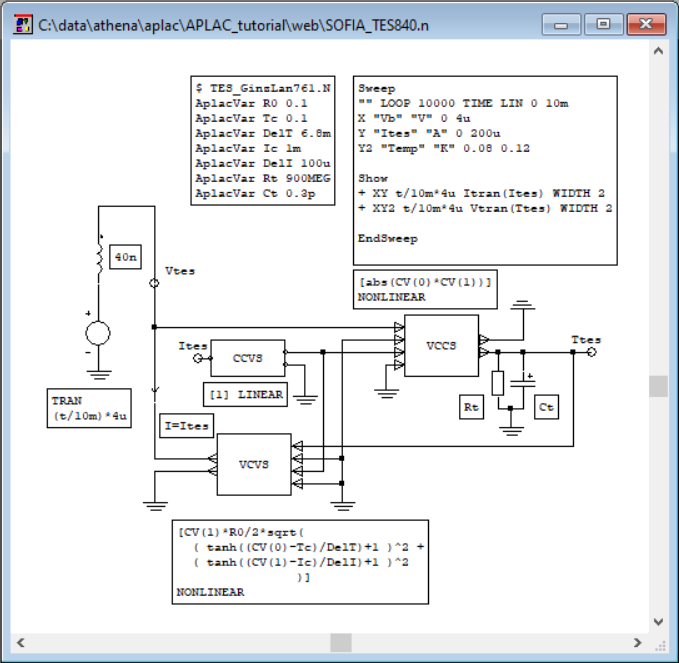}  \includegraphics[width=7.25cm]{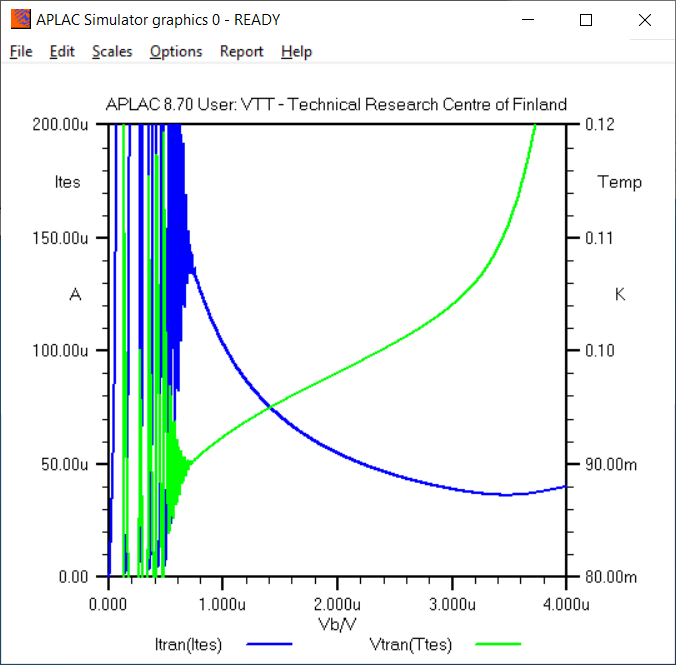}
\caption{A model for Transition Edge Sensors using controlled sources, including magnetically-induced switching from zero- to finite voltage state.} \label{TESmodel}
\end{figure}

Superconducting Transition Edge Sensors do not utilize the Josephson effect, but rely on a different non-linear phenomenon: the superconductive phase transition \cite{Irwin,WikiTES}. In the APLAC circuit of Fig.\ref{TESmodel}, we represent the internal temperature as a voltage across the thermal capacitance. The circuit model  \cite{SofiaTES} was developed in conjunction with the experiment \cite{KuurFdm}. Plotted is the TES current (blue) as a function of the bias voltage sweep, which shows the negative dynamic resistance region characteristic to TESes. The TES current shows the electrothermal oscillation at a low bias voltages owing to the inductance in series with the bias circuit. The internal TES temperature is plotted in green. The parameters: {\tt R0} is the normal-state resistance, {\tt Tc} the transition temperature, {\tt DelTwidth} of the thermal transition, {\tt Ic} critical current of the TES (necessary to get the simulation started) and {\tt DelI} width of the magnetic/current transition. {\tt Rt} is the thermal resistance to the bath and {\tt Ct} is the heat capacity. The APLAC schematic and netlist are available as {\tt SOFIA\_TES840.n} and {\tt SOFIA\_TES840.i}. 

The non-linear Voltage Controlled Voltace Source models the TES resistance in terms of the two control inputs {\tt CV(0)} and {\tt CV(1)}, representing the TES temperature and current, respectively. Because mixed-input controlled sources are not possible in APLAC, the linear Current Controlled Voltage Source is used to transform the TES current into a voltage that can be used for control purposes. The Voltage Controlled Current Source generates the heat flow driving the thermal capacitance {\tt Ct} as the product of TES current and TES voltage. 

\begin{figure}[h!] \centering
\includegraphics[width=7.25cm]{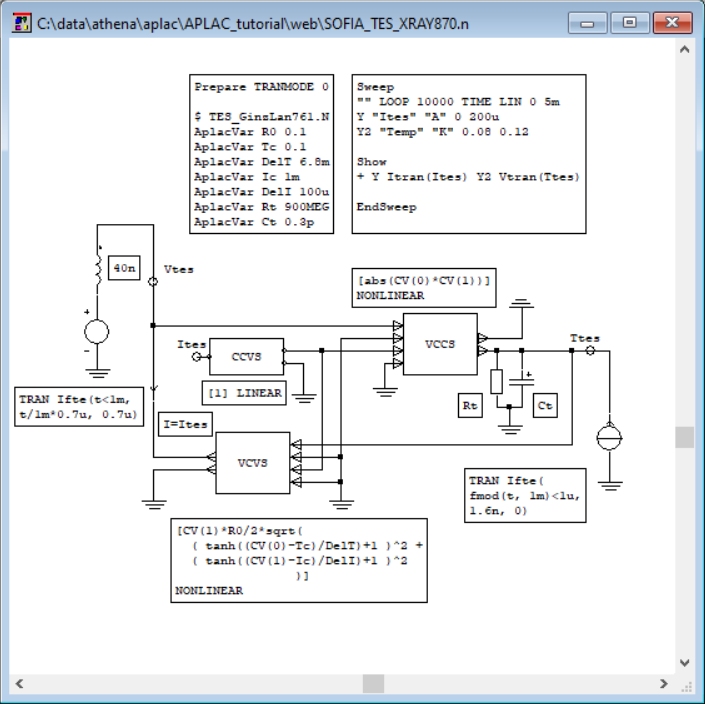}  \includegraphics[width=7.25cm]{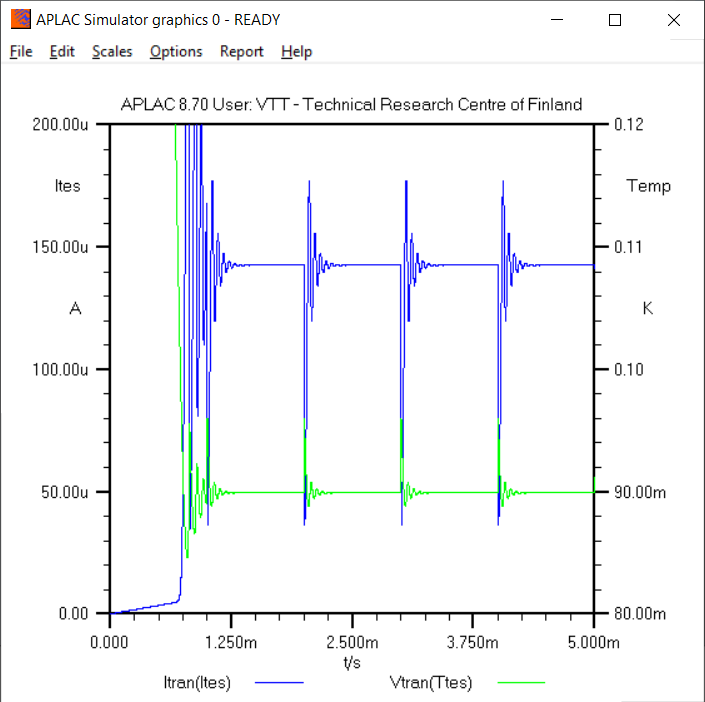}
\caption{The Transition Edge Sensor model driven with simulated X-ray photon absorption events.} \label{TES_xray}
\end{figure}

Additional heat flow (represented as APLAC current) can be arranged to drive the {\tt Ct} to simulate absorbed photons. In the circuit of Fig.\ref{TES_xray} the previous TES model is biased to the stationary \(U_{B} = 0.7 \:\mu V\), and arriving x-ray photons are simulated by \(1\mu s\) wide heat pulses whose area is adjusted to correspond the 10 keV photon energy. The {\tt Ifte} function (if-then-else) is used to drive the bias voltage from zero to its stationary value. The {\tt fmod} function (floating-point modulo) is used to arrange the regular x-ray photon arrival in every 1 ms. The TES current shows ringing, which implies that the particular TES still close to electrothermal instability at as low the bias voltage as the \(U_{B} = 0.7 \:\mu V\) than the shown dc case. In the picture where the voltage bias is visualized as a negative feedback servo, the electrothermal instability occurs when electric Joule feedback becomes slower that the thermal time constant of the TES. This can occur if the series inductance in the bias circuit is too large. The schematic and netlist are available as {\tt SOFIA\_TES\_XRAY870.n} and {\tt SOFIA\_TES\_XRAY870.i}.

\begin{figure}[h!] \centering
 \includegraphics[width=7.25cm]{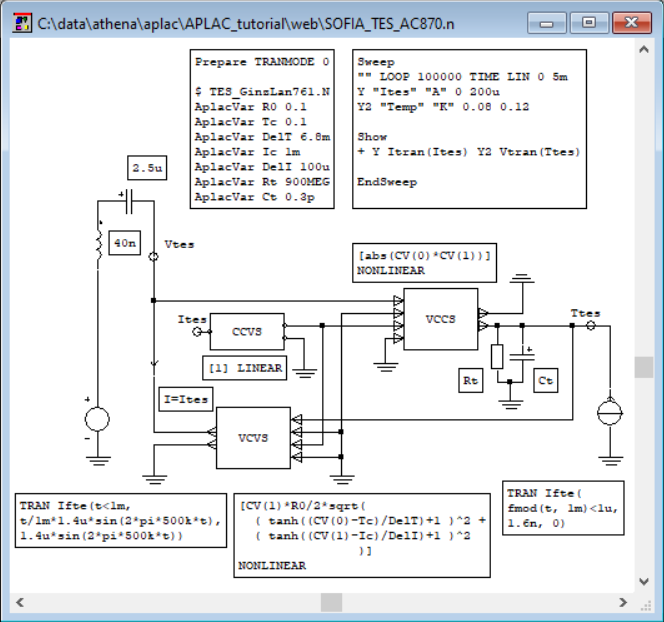}  \includegraphics[width=7.25cm]{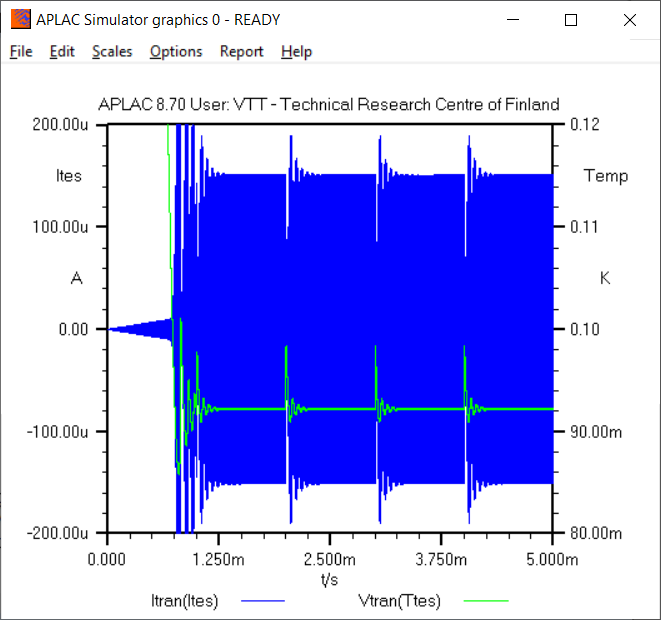}
 \includegraphics[width=14.5cm]{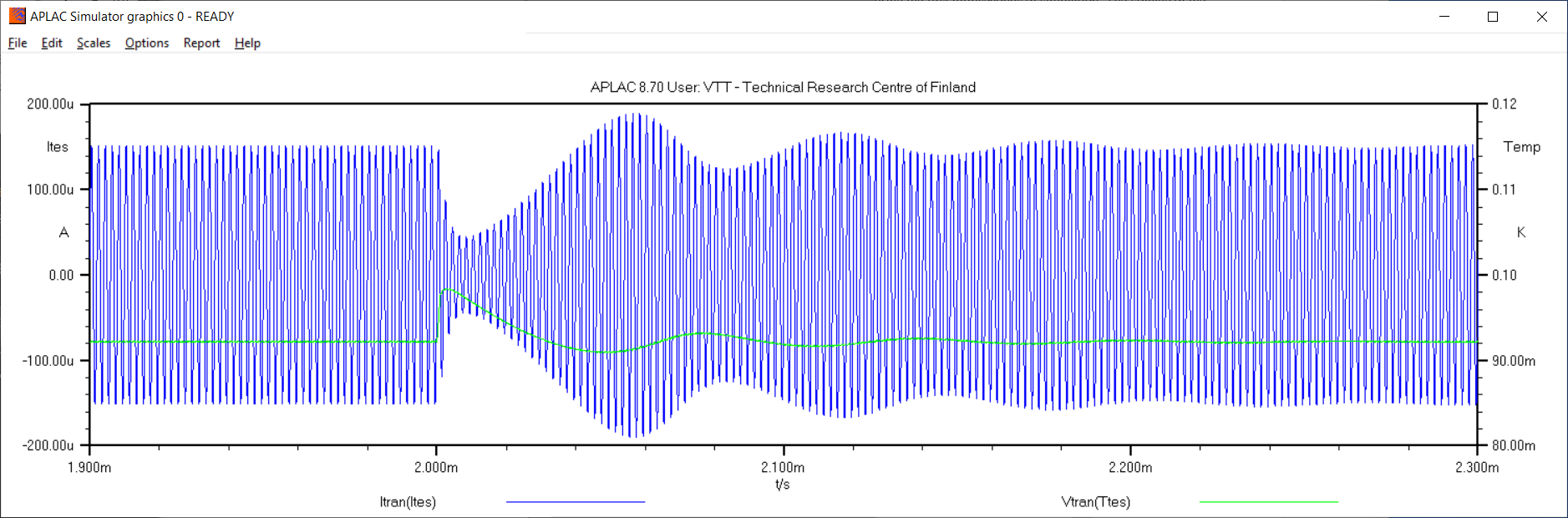}
\caption{Ac-biased Transition Edge Sensor modelled with APLAC, taken from \cite{SofiaTES} (upper). Magnified time trace of a photon absorption event (lower).} \label{TES_acbias}
\end{figure}

Of particular interest for us is the case of an AC biased TES, for implementaion of Frequency Domain Multiplexing. The TES in the Fig.\ref{TES_acbias} is equipped with an LC resonator, and is ac-biased at 500 kHz, with a slighly higher bias voltage \(U_{B} = 1 \:\mu V_{RMS}\). Regardless of higher bias, some electrothermal ripple is visible in the zoomed-in time trace, which shows the TES current after absorption of one x-ray photon. The marginal electrothermal stability is due to LC resonator settling time to be too slow relative to the thermal time constant, owing to too high Q-factor of the LC resonator. The APLAC {\tt SOFIA\_TES\_AC870.n} schematic and {\tt SOFIA\_TES\_AC870.i} netlist are available as ancilliary files.

\section{Summary}
We have found over the years APLAC to be a very useful tool in the design of practical SQUIDs and in understanding TES dynamics. Its long-standing commercial support and wide range of library components make it easier to simulate hybrids of Josephson-, TES- and more traditional electronic circuits. The APLAC facilities for hierarchical design alleviate modelling complex circuitry.

\end{document}